\documentclass[12pt]{article}
\usepackage{amssymb,graphicx,color,fancybox,epsfig,psfig}
\begin{document}

\baselineskip=.22in
\renewcommand{\baselinestretch}{1.2}
\renewcommand{\theequation}{\thesection.\arabic{equation}}

\begin{flushright}
{\tt hep-th/0510218 \\
KIAS-P05058 \\
SNUTP 05-015}
\end{flushright}

\vspace{5mm}

\begin{center}
{{\Large \bf DF-strings from D3${\bar {\bf D}}$3 as Cosmic Strings}\\[12mm]
Inyong Cho\\[2mm]
{\it Center for Theoretical Physics, School of Physics,\\
Seoul National University, Seoul 151-747, Korea}\\
{\tt iycho@phya.snu.ac.kr}\\[5mm]
Yoonbai Kim\\[2mm]
{\it BK21 Physics Research Division and Institute of Basic Science,}\\
{\it Sungkyunkwan University, Suwon 440-746, Korea}\\
{\tt yoonbai@skku.edu}\\[5mm]
Bumseok Kyae\\[2mm]
{\it School of Physics, Korea Institute for Advanced Study,\\
207-43, Cheongryangri-Dong, Dongdaemun-Gu, Seoul 130-012, Korea}\\
{\tt bkyae@kias.re.kr}
}
\end{center}

\vspace{5mm}

\begin{abstract}
We study Dirac-Born-Infeld type effective field theory of a
complex tachyon and U(1)$\times$U(1) gauge fields describing a
D3${\bar {\rm D}}$3 system. Classical solutions of straight global
and local DF-strings with quantized vorticity are found and are
classified into two types by the asymptotic behavior of the
tachyon amplitude. For sufficiently large radial distances, one
has linearly-growing tachyon amplitude and the other
logarithmically-growing tachyon amplitude. A constant radial
electric flux density denoting the fundamental-string background
makes the obtained DF-strings thick. The other electric flux
density parallel to the strings is localized, which represents
localization of fundamental strings in the D1-F1 bound states.
Since these DF-strings are formed in the coincidence limit of the
D3${\bar {\rm D}}$3, these cosmic DF-strings are safe from
inflation induced by the approach of the separated D3 and ${\bar
{\rm D}}3$.
\end{abstract}


\newpage

\setcounter{equation}{0}
\section{Introduction}
When we have a system of a D3-brane and an anti-D3-brane, its
dynamics is well described by the effective field theory of a
complex tachyon field and U(1)$\times$U(1) gauge
fields~\cite{Kraus:2000nj,Jones:2002si,Sen:2003tm}. While the D3
and ${\bar {\rm D}}3$ approach each other from apart, the Universe
undergoes an inflationary era due to the gravitational
effect~\cite{Dvali:1998pa}. When the D-brane coincides with the
${\bar {\rm D}}3$-brane, the system reaches the top of the tachyon
potential and the main inflation ends. Then, this unphysical
symmetric vacuum state at the zero tachyon amplitude restarts to
decay to the true U(1) degenerate vacua at an infinite tachyon
amplitude.

When the D3-brane and ${\bar {\rm D}}3$-brane are annihilated in
their coincidence limit, perturbative open string degrees living
on the branes disappear, but nonperturbative open string degrees
can survive in a form of fundamental strings, or of
lower-dimensional D-branes of codimension-two with closed string
degrees. In terms of effective field theory, one species among
those generated through a cosmological phase transition are
nothing but vortex-strings~\cite{VS,Kibble:2004hq} carrying D1-
(vorticity or quantized magnetic flux) and fundamental string
charge (electric flux along the string). Since inflation already
ended, the produced D1 and D1-F1 bound
states~\cite{Dvali:2003zj,Copeland:2003bj,Leblond:2004uc,
Blanco-Pillado:2005xx,Kim:2005tw,Polchinski:2004ia} can remain as
relics of the cosmic superstrings~\cite{Witten:1985fp} in the
present Universe.

In this paper we consider the DBI type effective action of a
complex tachyon and U(1)$\times$U(1) gauge fields, and find
straight global and local vortex-string solutions with an electric
flux. As shown in~\cite{Sen:2003tm,Kim:2005tw}, there exist static
global and local D-vortex solutions in the coincidence limit of
D2$\rm \bar{D}2$. While only singular D-vortex solutions are
possible without DBI electromagnetic fields~\cite{Sen:2003tm}, the
regular solutions are allowed when an electric flux is turned on
in the radial direction~\cite{Kim:2005tw}. The point-like
D-vortices could be readily extended to become D-strings of the
D3$\rm \bar{D}3$ system. In this paper we will also turn on a
constant electric flux along the string direction, and find that
its conjugate momentum density is well localized along the string.
The obtained static soliton configurations turn out to be
identified with DF-strings from a system of D3${\bar{\rm D}}3$
with fundamental string fluid. In addition to the known D-string
solutions with linearly-growing tachyon amplitude, we find new D-
and DF-string solutions with logarithmically-growing tachyon
amplitude.

Specific contents in each section are as follows. In section 2, we
introduce the effective action with a tachyon potential and
briefly discuss perturbative degrees and their fate on the
D$p{\bar {\rm D}}p$ system. In section 3, we first discuss global
DF-strings and then local DF-strings in details.  We conclude in section
4 with summary of the obtained results and discussions on a few
topics for future studies.

\setcounter{equation}{0}
\section{Setup and Perturbative Physics}
We consider a D$p{\bar {\rm D}}p$ system in the coincidence limit
of two branes, where the individual branes have the same
transverse coordinates. The brane-antibrane system possesses a
complex tachyon field $(T,\bar{T})$ describing instability of this
system and two massless gauge fields of U(1)$\times$U(1) symmetry
$A^{a}_{\mu},\;a=1,2$ living on each brane. Two representative
nonlocal effective actions have been used as tachyon actions,
i.e., one is derived from boundary string field theory
(BSFT)~\cite{Kraus:2000nj,Jones:2003ae} and the other is
Dirac-Born-Infeld (DBI) type proposed in Ref.~\cite{Sen:2003tm}.
In this paper we shall employ the latter,
\begin{eqnarray}\label{actl}
S=-{\cal T}_{p}\int d^{p+1}x\, V(\tau)\left[\,\sqrt{-\det(X^{+}_{\mu\nu})}
+\sqrt{-\det(X^{-}_{\mu\nu})}\,\,\right],
\end{eqnarray}
where ${\cal T}_{p}$ is tension of the D$p$-brane, $T=\tau e^{i\chi}$, and
\begin{equation}\label{Xpm}
X^{\pm}_{\mu\nu}=g_{\mu\nu}+F_{\mu\nu}\pm C_{\mu\nu}
+({\overline {D_{\mu}T}}D_{\nu}T +{\overline {D_{\nu}T}}D_{\mu}T)/2 .
\end{equation}
Our notations are
$A_{\mu}=(A^{1}_{\mu}+A^{2}_{\mu})/2$ with
$F_{\mu\nu}=\partial_{\mu}A_{\nu}-\partial_{\nu}A_{\mu}$,
$C_{\mu}=(A^{1}_{\mu}-A^{2}_{\mu})/2$ with
$C_{\mu\nu}=\partial_{\mu}C_{\nu}-\partial_{\nu}C_{\mu}$,
and $D_{\mu}T=(\partial_{\mu} -2iC_{\mu})T$.

Since DF-strings as codimension-two objects are of interest, we
consider D3${\bar {\rm D}}$3 system.   $-\det
(X^{\pm}_{\mu\nu})$ in the action with $p=3$ takes the following
form;
\begin{eqnarray}
-\det (X^{\pm}_{\mu\nu})&=&-\det
(g_{\mu\nu})\left[(1+S^\mu_\mu)\left(1+\frac{1}{2}{\cal
F}^{\pm}_{\rho\sigma}{\cal
F}^{\pm\rho\sigma}\right)+\frac{1}{2}A_{\mu\nu}A^{\mu\nu}+S^\mu_\nu
{\cal F}^{\pm}_{\mu\rho}{\cal F}^{\pm\rho\nu}\right]
\nonumber \\
&&-\frac{1}{64}\epsilon^{\mu\nu\rho\sigma}\epsilon^{\alpha\beta\gamma\delta}
{\cal F}^{\pm}_{\mu\nu}{\cal F}^{\pm}_{\rho\sigma}{\cal
F}^{\pm}_{\alpha\beta}{\cal F}^{\pm}_{\gamma\delta}-\frac{1}{16}
\epsilon^{\mu\nu\rho\sigma}\epsilon^{\alpha\beta\gamma\delta}
{\cal F}^{\pm}_{\mu\nu}A_{\rho\sigma}{\cal
F}^{\pm}_{\alpha\beta}A_{\gamma\delta}\, ,
\label{xde}
\end{eqnarray}
where ${\cal F}^{\pm}_{\mu\nu}\equiv F_{\mu\nu}\pm C_{\mu\nu}$,
$S_{\mu\nu}\equiv (\overline{D_\mu T}D_\nu T+\overline{D_\nu
T}D_\mu T)/2$, and $A_{\mu\nu}\equiv (\overline{D_\mu T}D_\nu
T-\overline{D_\nu T}D_\mu T)/2i$, respectively. Up to the
quadratic terms in the gauge fields and derivative of the tachyon
amplitude, the Lagrange density in (1+3) dimensions becomes
\begin{eqnarray}\label{la}
{\cal L}\approx -2{\cal T}_{3}V(\tau)\left[\left(\frac{1}{2}\partial_{\mu}\tau
\partial^{\mu}\tau+1\right)+\frac{1}{4}F_{\mu\nu}F^{\mu\nu}
+\left(\frac{1}{4}C_{\mu\nu}C^{\mu\nu}+2\tau^{2}\tilde{C}_{\mu}\tilde{C}^{\mu}
\right) \right] ,
\end{eqnarray}
where the unitary gauge,
$\tilde{C}_{\mu}=C_{\mu}-\partial_{\mu}\Omega/2$, is chosen for
topologically trivial sector with
$C_{\mu\nu}=\partial_{\mu}\tilde{C}_{\nu}-\partial_{\nu}\tilde{C}_{\mu}$.
Note that a cross term $F_{\mu\nu}C^{\mu\nu}$ does not appear in
the approximated Lagrange density (\ref{la}).

From Ref.~\cite{Sen:1999mg,Sen:1999xm}, universally allowed
conditions of the tachyon potential $V$ for the ${\rm D}p\bar{{\rm
D}}p$ system are monotonically decaying property connecting
smoothly the maximum of $V(\tau=0)=1$ and minimum of
$V(\tau=\infty)=0$.  To support perturbative spectrum in
superstring theory, we choose
$d^{2}V/d\tau^{2}|_{\tau=0}=-1/R^{2}=-1/2$. In the DBI type
effective action, exponentially decaying property for large
$\tau$, $V(\tau)\sim e^{-\tau/R}$ is usually
assumed~\cite{Sen:2002an}. For the analysis of DF-string solutions
with the cylindrical symmetry, the above properties are enough at
both string core and asymptotic region.  For numerical analysis,
however, we will use a specific potential satisfying all the above
conditions for convenience~\cite{Kutasov:2003er,Buchel:2002tj}
\begin{equation}\label{V3}
V(\tau)=\frac{1}{{\rm cosh}\left(\frac{\tau}{R}\right)}.
\end{equation}
Let us read possible perturbative spectra from the Lagrange
density (\ref{la}). Before the ${\rm D}3\bar{{\rm D}}3$ decays,
the complex scalar fields, $T$ and $\bar{T}$, are tachyonic, and
both gauge fields, $A_{\mu}$ and $C_{\mu}$, are massless. When it
decays completely, a ring of degenerate minima at infinite tachyon
amplitude is formed. Naively $A_{\mu}$ seems to remain massless
and $\tilde{C}_{\mu}$, absorbing the Goldstone degree $\Omega$,
becomes massive due to nonzero vacuum expectation value of $\tau$.
Different from usual field theory results, all the tachyon and the
gauge fields cannot survive due to vanishing tachyon potential
which is an overall factor in the Lagrange density (\ref{la}).
This phenomenon is easily expected because all the perturbative
open string degrees should disappear after the decay of ${\rm
D}3\bar{{\rm D}}3$. On the other hand, nonperturbative degrees
including codimension-two branes and fundamental strings can be
formed, so that the runaway nature of above tachyon potential
should play an indispensable role for determining characters of
the generated topological solitons.

\setcounter{equation}{0}
\section{DF-strings}

In this section we study static DF-string solutions of the
classical equations, which are identified as the codimension-two
DF-composites from D3${\bar {\rm D}}$3. The obtained nonsingular
configurations are classified into the following four types by two
crossed borderlines, i.e., (i) global U(1) DF-vortices with
critical boundary value of electric field at infinity
$F_{tr}(r=\infty)$, (ii) global U(1) DF-vortices with noncritical
boundary values of $F_{tr}(r=\infty)$, (iii) local U(1)
DF-vortices with critical boundary value of $F_{tr}(r=\infty)$,
and (iv) local U(1) DF-vortices with noncritical boundary values
of $F_{tr}(r=\infty)$. Since there is one-to-one correspondence
between the obtained DF-vortex solution and the D-vortex solution
in Ref.~\cite{Kim:2005tw}, the newly-obtained DF-vortices with
critical boundary value imply the existence of additional
D-vortices with the same critical electric field.

Straight strings along the $z$-axis is conveniently described in
the cylindrical coordinates $(t,r,\theta,z)$.  The ansatz for the
$n$ strings superimposed at the origin $r=0$ is
\begin{equation}\label{ant}
T=\tau(r) e^{in\theta}.
\end{equation}
In order to obtain regular DF-strings, we assume the minimal
configuration of the DBI electromagnetic fields $F_{\mu\nu}$ as
\begin{equation}\label{ane}
F_{tr}(r)=E_{r}(r),\quad F_{tz}(r)=E_{z}(r),\quad \mbox{others}=0.
\end{equation}
Introduction of the gauge field $C_{\theta}$ replaces global
strings by local strings
\begin{eqnarray}\label{anc}
C_{\mu}=\delta_{\mu \theta}C_{\theta}(r), \quad (C_{r\theta}=C_{\theta}').
\end{eqnarray}

Insertion of the ans\"{a}tze (\ref{ant})--(\ref{anc}) into the
determinant (\ref{xde}) gives
\begin{eqnarray} \label{det}
&&-\det (X^{\pm}_{\mu\nu})
\equiv  r^2X \\
&&=-\det
(g_{\mu\nu})\left\{\left[1+\frac{\tau^2}{r^2}(n-2C_{\theta})^2\right]\left[
\left(1-E_{z}^2\right) \left(1+{\tau'}^2\right) -E_{r}^2 \right]
+\left(1-E_{z}^2\right)\frac{{C_{\theta}'}^2}{r^2}\right\} ,
\label{det1}
\end{eqnarray}
which simplifies the action (\ref{actl}) as
\begin{equation}
S=-2{\cal T}_{3}\int dt dr d\theta dz r\, V(\tau) \sqrt{X}\, .
\end{equation}

Bianchi identity, $\partial_\mu F_{\nu\rho}+\partial_\nu
F_{\rho\mu}+\partial_\rho F_{\mu\nu}=0$, requires $E_z$ to be a
constant. When $E_{z}^{2}>1$, $X$ becomes negative and the action
(\ref{actl}) becomes imaginary, which is physically unacceptable.
When $E_{z}^{2}=1$, derivative of the tachyon amplitude disappears
in (\ref{det1}) and then no nontrivial solution is supported. When
$E_{z}^{2}<1$, introduction of new variables,
\begin{eqnarray}\label{corr}
{\tilde E}_{r}(r)=\frac{E_{r}}{\sqrt{1-E_{z}^2}},\qquad
{\tilde {\cal T}}_{3}=\sqrt{1-E_{z}^2}\, {\cal T}_{3},\qquad
\tilde{X}=\frac{X}{1-E_{z}^2},
\end{eqnarray}
show that we have $\tilde{X}=X|_{E_z=0}$ and thereby
the system with nonvanishing constant $E_{z}$ is formally equivalent to
that with vanishing $E_{z}$ under the correspondence (\ref{corr}).

For the gauge field $A_{i}$ and conjugate momentum $\Pi^{i}$,
the only nontrivial equation is
$(r\Pi^r)'=0$ which is rewritten by introducing constant charge
density $Q_{{\rm F}1}$ per unit length along $z$-axis as
\begin{equation}\label{Pir}
\Pi^{r}\equiv \frac{1}{\sqrt{-g}}\frac{\delta S}{\delta
(\partial_{t}A_{r})} =\frac{1}{\sqrt{-g}}\frac{\delta S}{\delta
E_{r}}= \frac{1}{\sqrt{1-E_{z}^2}}\frac{2{\tilde {\cal
T}}_3V}{\sqrt{\tilde{X}}}\left[1+\frac{\tau^2}{r^2}(n-2C_\theta)^2\right]
{\tilde E}_r =\frac{Q_{{\rm F}1}}{r}.
\end{equation}
Equation of motion for the tachyon amplitude $\tau$ is
\begin{eqnarray} \label{Teq}
\frac{1}{r}\frac{d}{dr}\left\{
\frac{rV}{\sqrt{\tilde{X}}}\left[1+
\frac{\tau^2}{r^2}(n-2C_\theta)^2\right]\tau' \right\}
-\frac{V}{\sqrt{\tilde{X}}}\left(
1+\tau^{'2} -\tilde{E}_{r}^2
\right)\frac{(n-2C_\theta)^2}{r^{2}}\tau
=\sqrt{\tilde{X}}\frac{dV}{d\tau} ,
\end{eqnarray}
and that for the gauge field $C_{\theta}$ is
\begin{eqnarray} \label{Ceq}
\frac{1}{r}\frac{d}{dr}\left( \frac{rV}{\sqrt{\tilde{X}}}
\frac{{C_{\theta}'}}{r^2}\right)
+2\frac{V}{\sqrt{\tilde{X}}}\left( 1+\tau^{'2}-\tilde{E}_{r}^2
\right)\frac{\tau^2(n-2C_\theta)}{r^{2}}=0 .
\end{eqnarray}
From (\ref{Pir}) we obtain an algebraic expression for $E_r$ (or equivalently
${\tilde E}_{r}$)
\begin{eqnarray}\label{Er}
{\tilde E}_{r}(r)^{2}=\frac{E_r(r)^{2}}{1-E_z^2} =\frac{(1+\tau^{'2})
\left[1+\frac{\tau^2}{r^2}(n-2C_\theta)^2\right]
+\frac{C_\theta^{'2}}{r^2}}
{\left[1+\frac{\tau^2}{r^2}(n-2C_\theta)^2\right]
\left\{1+\left(\frac{2{\cal T}_3 rV}{Q_{{\rm F}1}}\right)^2
\left[1+\frac{\tau^2}{r^2}(n-2C_\theta)^2\right]\right\}} .
\end{eqnarray}

The $\tau$- and $C_\theta$-equations (\ref{Teq})--(\ref{Ceq}) with
$E_z\ne 0$ is exactly the same as the equations with $E_z=0$. The
solutions have been discussed in Ref.~\cite{Kim:2005tw}
\begin{equation}
\tau (r) =\tau (r) |_{E_z=0} ,\qquad
C_\theta(r) = C_\theta (r)|_{E_z=0} .
\end{equation}
The functional form of $E_{r}(r)|_{E_z=0}$ has been also discussed
in~\cite{Kim:2005tw}, which is the same as $E_r(r)$ in (\ref{Er})
except for an overall factor $(1-E_z^2)$. According to
Ref.~\cite{Kim:2005tw}, the obtained D-vortex solutions are
classified as follows. With nonvanishing $E_{r}$ regular vortex
solutions are obtained, but with vanishing
$E_{r}$ only singular configurations are
constructed~\cite{Sen:2003tm}. Characters of the obtained
vortices are divided by the gauge field $C_{\mu}$, i.e., global
vortices for $C_{\mu}=0$ and local vortices for $C_{\mu}\ne 0$.
Since the extension from the point-like D-vortices to the straight
D-strings along $z$-direction is straightforward, the
aforementioned properties of D-vortex solutions hold also for the
DF-strings of our interest.

In the above we have explained similarity between the vortex
solutions without $E_{z}$ and those with $E_{z}$. Let us discuss
the quantities how to distinguish DF-vortices with $E_{z}$ from
D-vortices without $E_{z}$ in what follows. Once we obtain
profiles of the tachyon amplitude $\tau$ and the gauge field
$C_{\theta}$ for given constant $Q_{{\rm F}1}$ and $E_{z}$, the
fundamental string charge density per unit length distributed
along the straight DF-string is given by conjugate momentum
$\Pi^z$ of the gauge field $A_{z}$
\begin{eqnarray}
\Pi^z(r)^2&\equiv& \left[\frac{1}{\sqrt{-g}}\frac{\delta S}{\delta
(\partial_{t}A_{z})}\right]^2=\left(\frac{1}{r}\frac{\delta
S}{\delta E_{z}}\right)^2
\nonumber\\
&&\hspace{-22mm}=\frac{Q_{{\rm F}1}^{2}E_z^2}{1-E_z^2} \,
\frac{(1+\tau^{'2})
\left[1+\frac{\tau^2}{r^2}(n-2C_\theta)^2\right]
+\frac{C_\theta^{'2}}{r^2}}{r^2\left[1+\frac{\tau^2}{r^2}(n-2C_\theta)^2
\right]} \,
\left\{1+\left(\frac{2{\cal T }_3rV}{Q_{{\rm F}1}}\right)^2
\left[1+\frac{\tau^2}{r^2}(n-2C_\theta)^2\right]\right\} .
\label{Piz}
\end{eqnarray}
To be identified as a DF-string, $\Pi^z$ should be localized on
the D-string stretched along $z$-direction. Although the shape of
the fundamental string charge density per unit length keeps the
same form (\ref{Pir}) irrespective of its charge $Q_{{\rm F}1}$,
that of the DF-strings changes its shape by $Q_{{\rm F}1}$.

To understand detailed property of the DF-strings we also need to
investigate U(1) current $j^{\theta}$
\begin{eqnarray}\label{jth}
j^\theta = \frac{2{\tilde {\cal T}}_3V}{\sqrt{{\tilde X}}}\left(
1+\tau^{'2}-{\tilde E}_r^2\right) \frac{\tau^2}{r^2}(n-2C_\theta),
\end{eqnarray}
and nonvanishing components of the energy-momentum tensor
\begin{eqnarray}
T^t_{\; t} &=& -\frac{2{\tilde {\cal T}}_3V}{(1-E_z^2)\sqrt{{\tilde
X}}} \left\{\left[1+\frac{\tau^2}{r^2}(n-2C_\theta)^2\right]
(1+\tau^{'2})+\frac{C_\theta^{'2}}{r^2}\right\} ,
\label{Tt}\\
T^r_{\; r} &=& -\frac{2{\tilde {\cal T}}_3V}{
\sqrt{{\tilde X}}}\left[1+\frac{\tau^2}{r^2}(n-2C_\theta)^2\right],
\label{Tr}\\
T^\theta_{\;\theta} &=& -\frac{2{\tilde {\cal T}}_3V}{
\sqrt{{\tilde X}}}\left(1+\tau^{'2}-{\tilde E}_r^2\right) ,
\label{Tth}\\
T^z_{\; z} &=& -\frac{2\tilde {\cal T}_3V}{(1-E_z^2)\sqrt{ \tilde
X}}\left\{\left[1+\frac{\tau^2}{r^2}(n-2C_\theta)^2\right]
(1+\tau^{'2}-E_r^2)+\frac{C_\theta^{'2}}{r^2}\right\} .
\label{Tz}
\end{eqnarray}

\subsection{Global DF-strings}

Global DF-vortex solutions are attained by choosing constant gauge field
$C_{\theta}=0$ in the previous part of the section 2 with neglecting
the gauge field equation (\ref{Ceq}).
Then the only nontrivial differential
equation is that of the tachyon amplitude (\ref{Teq}).
For every regular vortex solution of $n\ne 0$,
boundary conditions for the tachyon amplitude are
\begin{equation}\label{bd}
\tau(r=0)=0,\qquad \tau(r\rightarrow \infty)\rightarrow \infty.
\end{equation}
The runaway nature of the tachyon potential dictates that
$\tau(r)$ of a DF-string solution should be a
monotonically-increasing function which connects smoothly the
boundaries with the conditions (\ref{bd}).

Near the origin, a consistent power-series expansion leads to
increasing $\tau$,
\begin{equation}\label{tr00}
\tau(r)\approx
\left\{
\begin{array}{ll}
\tau_0 r \left[1 -\frac{{\cal T}_3^2(1+\tau_0^2)^2}{5 Q_{\rm F1}^2R^2}r^4
+ \cdots\right], & (n=1), \\
\tau_{0}r\left[1+\frac{2{\cal T}_3^{2}}{3Q_{{\rm
F}1}^{2}}(1+\tau_0^2)(n^2-1)r^2 - {\cal O}(r^4)\right], & (n \ge 2).
\end{array}
\right.
\end{equation}
Inserting (\ref{tr00}) into (\ref{Er}) and (\ref{Piz}),
we have decreasing $E_{r}^{2}$ from a constant value and decreasing
$\Pi^{z2}$ from the infinity,
\begin{eqnarray}
E_{r}^{2} &\approx & (1-E_{z}^{2}) (1+\tau_0^2)\left[
1-\frac{4{\cal T}_p^{2}}{Q_{{\rm F}1}^{2}}(1+\tau_0^2) r^2 +\cdots
\right]
, \label{ee0}\\
\Pi^{z2}&\approx & \frac{E_z^2(1+\tau_0^2)}{1-E_z^2}
\left(\frac{Q_{{\rm F}1}}{r}\right)^{2}\left\{1 +\frac{4{\cal T}_3^2}{Q_{\rm
F1}^2}\left[1+\tau_0^2(2n^2-1)\right]r^2 +\cdots\right\} .
\end{eqnarray}
In addition, we obtain the current density (\ref{jth}),
\begin{eqnarray}\label{jth0}
j^{\theta}&\approx & 4n\tilde{{\cal
T}}_3^2\tau_0^2 \sqrt{\frac{1+\tau_0^2}{1-E_z^2}}~
\frac{r}{|Q_{\rm F1}|} + \cdots ,
\end{eqnarray}
and the energy-momentum tensor (\ref{Tt})--(\ref{Tz}),
\begin{eqnarray}
T^{t}_{\; t}&\approx &
-\sqrt{\frac{1+\tau_0^2}{1-E_z^2}}~\frac{|Q_{\rm F1}|}{r}
\left\{1+\frac{2{\cal T}_3^2}{Q_{\rm F1}^2}\left[1+\tau_0^2(2n^2-1)\right]r^2
+\cdots\right\} ,
\label{tt0}\\
T^{r}_{\; r}&\approx &
-\sqrt{\frac{1-E_z^2}{1+\tau_0^2}}~\frac{|Q_{\rm F1}|}{r}
\left[1+\frac{2{\cal T}_3^2}{Q_{\rm F1}^2}(1+\tau_0^2)r^2 +\cdots\right] ,\\
T^{\theta}_{\; \theta}&\approx & -4{\tilde{\cal
T}}_3^2\sqrt{\frac{1+\tau_0^2}{1-E_z^2}}
~\frac{r}{|Q_{\rm F1}|} + \cdots ,
\label{tth0} \\
T^{z}_{\; z}&\approx &
-\sqrt{\frac{1+\tau_0^2}{1-E_z^2}}~\frac{|Q_{\rm
F1}|}{r}\left\{E_z^2 + \frac{2{\cal T}_3^2}{Q_{\rm F1
}^2}[2(1+\tau_0^2n^2)-E_z^2(1+\tau_0^2)]r^2 +\cdots \right\} .
\label{tz0}
\end{eqnarray}
As it was expected, the angular component of U(1) current $j^{\theta}$ and the
pressure $T^{\theta}_{\theta}$ vanish at the origin.
In $T^{t}_{t}$, $T^{r}_{r}$, and $T^{z}_{z}$, the leading term shows
singular behavior due to the background fundamental string charge
$Q_{{\rm F1}}$.
As this fundamental-string charge decreases to zero, the leading term
goes to zero, but the slope of the second term proportional to $1/Q_{{\rm F}1}$
becomes steep. It is consistent with the observation that only the singular
global vortex solution exists in the absence of
the background fundamental-string
charge~\cite{Kim:2005tw}.

At sufficiently large $r$, we solve the tachyon equation
(\ref{Teq}).  We try to get the tachyon solutions with (i)
power-law behavior $\tau\sim \tau_\infty r^k$ ($k> 0$) and (ii)
logarithmic behavior $\tau\sim \ln r$.

\begin{figure}[t]
\centerline{\epsfig{figure=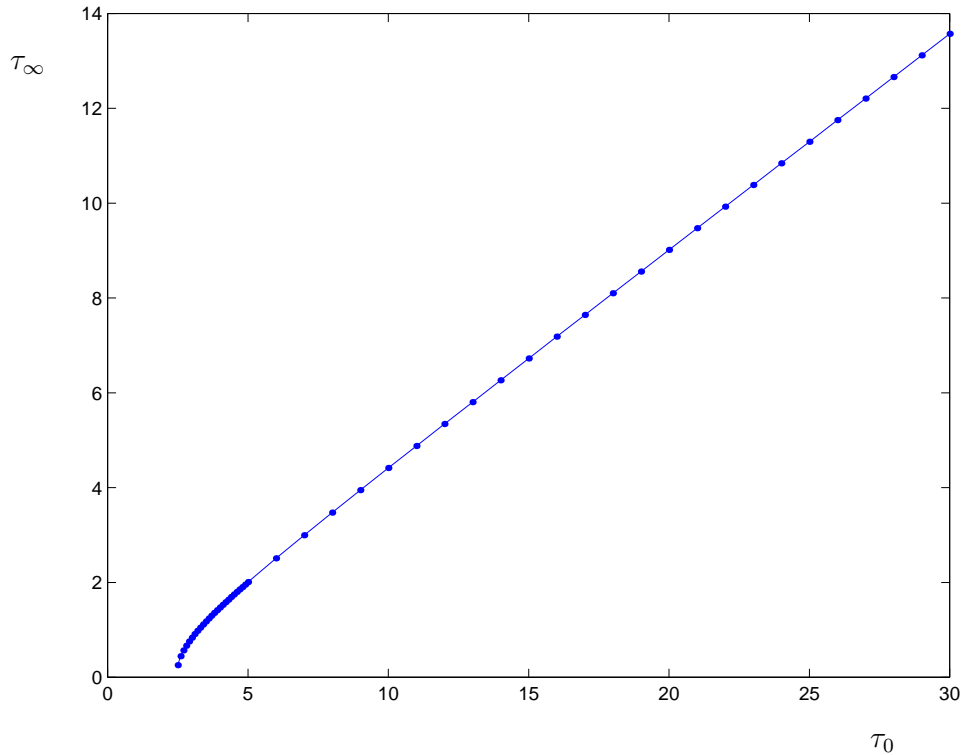,height=100mm}}
\caption{Plot of
$\tau_0$ vs. $\tau_\infty$ for $n=1$ DF-vortex with $Q_{{\rm
F1}}/{\cal T}_{3}=2$ and $R=\sqrt{2}$. At $\tau_{0}=2.46327$,
$\tau_{\infty}\rightarrow 0$ which supports the
logarithmically-increasing $\tau(r)$ at asymptotic region.}
\label{fig1}
\end{figure}

{\bf (i) \underline{$\tau\sim \tau_\infty r$ solutions}}: If we
substitute the power-law behavior into the tachyon equation
(\ref{Teq}), only the linearly increasing $\tau$ solution is
allowed at leading order. The power-series expansion gives
\begin{eqnarray} \label{solT}
\tau &\approx& \tau_\infty r + \delta  - \frac{4{\cal T
}_3^2R}{\tau_\infty^2Q_{\rm
F1}^2}(1+\tau_\infty^2)(1+\tau_\infty^2n^2)
r^2e^{-2\frac{\tau_\infty r+\delta}{R}} + \cdots ,
\end{eqnarray}
where $\tau_\infty$ ($>0$) and $\delta$ are undetermined, but
$\tau_\infty$ is related with $\tau_{0}$ near the origin.
Numerical works show that $\tau_{\infty}$ is almost proportional
to $\tau_{0}$ for large $\tau_{0}$'s as in Fig.~\ref{fig1}.
Specifically, for $n=1$, $Q_{{\rm F1}}/{\cal T}_{3}=2$, and $R=\sqrt{2}$,
$\displaystyle{\lim_{\tau_{0}\rightarrow \infty}(\tau_{\infty}/
\tau_{0})\rightarrow 0.4566}$.

\begin{figure}[t]
\centerline{\epsfig{figure=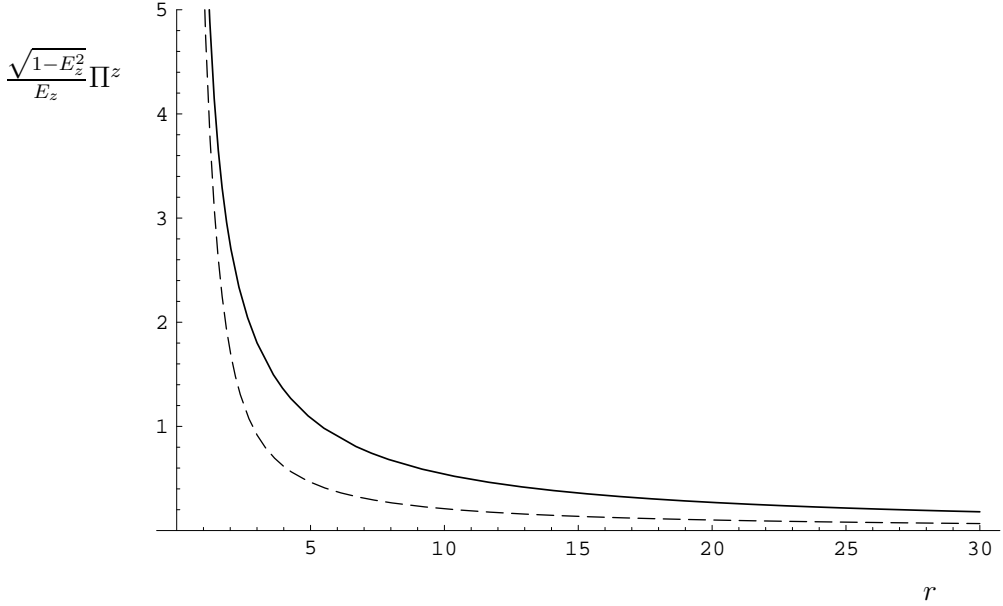,height=80mm}}
\caption{ Plot of
$\Pi^z(r)$ for the asymptotically-logarithmic (dashed line:
$\tau_{0}=2.46327$) and asymptotically-linear (solid line:
$\tau_{0}=6$) configurations of $\tau$. We set $n=1$, $Q_{{\rm
F1}}/{\cal T}_{3}=2$, and $R=\sqrt{2}$ for superstring theory. }
\label{fig2}
\end{figure}

Inserting (\ref{solT}) into the various physical quantities
(\ref{Er}), (\ref{Piz}), (\ref{jth})--(\ref{Tz}), we read the
followings. First, the radial electric field approaches rapidly a
constant boundary value
$|E_{r}(\infty)|=\sqrt{(1-E_{z}^{2})(1+\tau_{\infty}^{2})}\,$ ,
\begin{eqnarray}\label{solE}
E_r^2&\approx& (1-E_{z}^{2})
(1+\tau_\infty^2)\left\{1-\frac{16{\cal T}_3^2R}{\tau_\infty
Q_{\rm
F1}^2}\left[1+\tau_\infty^2n^2\left(1+\frac{2\delta}{R}\right)
\right]re^{-2\frac{\tau_\infty r+\delta}{R}} + \cdots\right\}  .
\end{eqnarray}
Second, the leading terms of $\Pi^{z}$, $T^{t}_{t}$, $T^{r}_{r}$,
$T^{z}_{z}$ are all proportional to $\Pi^{r}$ ($=Q_{{\rm F}1}/r$)
and the subleading terms exponentially suppressed,
\begin{eqnarray}
\Pi^{z2}&\approx &  \frac{E_z^2(1+\tau_\infty^2)}{1-E_z^2}
\left(\frac{Q_{\rm F1}}{r}\right)^{2} \left[1 + \frac{32{\cal
T}_3^2}{Q_{\rm F1}^2}(1+\tau_\infty^2n^2)r^2
e^{-2\frac{\tau_\infty r+\delta}{R}}
+\cdots \right] ,
\label{pzz}\\
T^{t}_{\; t}&\approx & -\sqrt{\frac{1+\tau_{\infty}^{2}}{1-E_z^2}}
\,~\frac{|Q_{{\rm F1}}|}{r}\left[1+\frac{16{\cal T }_3^2}{Q_{\rm
F1}^2}(1+\tau_\infty^2n^2)r^2e^{-2\frac{\tau_\infty
r+\delta}{R}}+ \cdots \right],
\label{ett2}\\
T^{r}_{\; r}&\approx & -\sqrt{\frac{1-E_z^2}{1+\tau_{\infty}^{2} }}
\,~\frac{|Q_{{\rm F1}}|}{r}\left\{1+\frac{8{\cal T
}_3^2R}{\tau_\infty Q_{\rm
F1}^2}\left[1+\tau_\infty^2n^2\left(1+\frac{2\delta}{R}\right)\right]
re^{-2\frac{\tau_\infty r+\delta}{R}} +\cdots\right\},\\
T^{z}_{\; z}&\approx &
-\sqrt{\frac{1+\tau_\infty^2}{1-E_z^2}}~\frac{|Q_{\rm
F1}|}{r}\left[E_z^2 + \frac{16{\cal T}_3^2}{Q_{\rm
F1}^2}(1+\tau_\infty^2n^2)r^2e^{-2\frac{\tau_\infty
r+\delta}{R}}+\cdots \right] .
\label{tzu}
\end{eqnarray}
Third, the angular components $j^{\theta}$ and
$T^{\theta}_{\theta}$ exponentially decreasing so that, with
(\ref{jth0}) and (\ref{tth0}), their shapes in $(r,\theta)$-plane
look like a ring,
\begin{eqnarray}
j^{\theta}&\approx & 16n\tilde{{\cal
T}}_{3}^2\tau_{\infty}^2\sqrt{\frac{1+\tau_{\infty}^{2}}{1-E_z^2}
}~\frac{|Q_{\rm F1}|}{r} e^{-2\frac{\tau_\infty r+\delta}{R}}+\cdots ,
\label{jtt}\\
T^{\theta}_{\;\theta}&\approx & -16\tilde{{\cal
T}}_{3}^2\sqrt{\frac{1+\tau_{\infty}^{2}}{1-E_z^2}}~\frac{r}{|Q_{\rm
F1}|}e^{-2\frac{\tau_\infty r+\delta}{R}}+\cdots .
\label{tthu}
\end{eqnarray}
Here, we do not present the results of numerical analysis since
the obtained configurations are exactly the same as those of
D-strings~\cite{Kim:2005tw} except for the fundamental-string
charge density $\Pi^{z}$ in Fig.~\ref{fig2}.

{\bf (ii) \underline{$\tau\sim \tau_\infty \ln r$ solution}}: If
$\tau_{0}$ in the tachyon field near the origin (\ref{tr00}) is
sufficiently small, then this solution cannot reach
$\tau(r=\infty)=\infty$. It means that there exists a critical
value of $\tau_{0}$ which corresponds to $\tau'(\infty
)\rightarrow 0$ in (\ref{solT}), and it also requires a
critical-charge density $(Q_{{\rm F}1}/{\cal T}_{3})^{2}=
8/R^{2}$. In this limit, a natural asymptotic behavior of the
tachyon amplitude is logarithmic, $\tau(r)\sim \tau_{\infty}{\rm
ln}r$.\footnote{In the context of 4 dimensional supergravity, a
logarithm-type behavior at asymptotic region was
found in the dilaton field ($\sim \ln r$) and the Higgs field
($\sim {\rm VEV}-{\cal O}(1/{\rm ln} r)$), and the obtained vortices
carry finite tension~\cite{Blanco-Pillado:2005xx}.} If we
try this configuration, the field equation (\ref{Teq}) with
(\ref{Er}) fixes the value of $\tau_{\infty}$ to
$\tau_{\infty}=2R$, which leads the tachyon potential to a
power-law decay, $V\approx 2/r^{2}$;
\begin{eqnarray} \label{logsolT}
\tau(r)\approx  2R\, {\rm ln}r\left(1 -2n^2R^2\frac{{\rm
ln}r}{r^2} + \cdots\right) .
\end{eqnarray}
Note that $\ln r$ is not well-defined at the entire region $(0\le
r\le \infty)$, regularity of the obtained solution needs further
mathematically-rigorous study. It turns out, in this case, that
the radial component of the electric field $E_{r}$ (\ref{Er})
approaches a critical value at infinity with a power-law ${\cal
O}(1/r^{2})$, $E_{r}^2(r=\infty)=1-E_{z}^{2}$,
\begin{eqnarray} \label{loer}
E_{r}^{2}(r)\approx  (1-E_z^2)\left(1+ \frac{2R^{2}}{r^{2}}
+ \cdots \right) .
\end{eqnarray}
This looks similar to the case of the thick single topological BPS tachyon
kink~\cite{Kim:2003in}.

Inserting (\ref{logsolT}) into (\ref{Piz}), (\ref{Tt}), (\ref{Tr}),
and (\ref{Tz}), we have again ${\cal O}(1/r)$ leading term in
$\Pi^{z}$, $T^{t}_{\; t}$, $T^{r}_{\; r}$, and $T^{z}_{z}$,
\begin{eqnarray}
\Pi^{z2}&\approx & \frac{E_z^2}{1-E_z^2}\left(\frac{Q_{\rm
F1}}{r}\right)^2\left(1+\frac{6R^2}{r^2}
+\cdots \right) ,
\label{logsolE}\\
T^{t}_{\; t}&\approx & -\frac{1}{\sqrt{1-E_z^2}}
\,\frac{|Q_{\rm F1}|}{r}\left(1+\frac{3R^2}{r^2}+\cdots\right),
\label{trin}\\
T^{r}_{\; r} &\approx & -\sqrt{1-E_z^2} \,\frac{|Q_{\rm
F1}|}{r}\left(1-\frac{R^2}{r^2}+\cdots\right),
\\
T^{z}_{z}&\approx & -\frac{1}{\sqrt{1-E_z^2}}\frac{|Q_{\rm
F1}|}{r}\left[E_z^2 +(2+E_z^2)\frac{R^2}{r^2}+\cdots \right].
\label{ltz}
\end{eqnarray}
The coefficients of the leading terms can be understood as the $\tau_{\infty}
\rightarrow 0$ limit of (\ref{pzz})--(\ref{tzu}) for the power-law
solution.
However, the subleading terms exhibit also a power-law behavior in
(\ref{logsolE})--(\ref{ltz}), instead of the exponential decay in
(\ref{pzz})--(\ref{tzu}).
This ${\cal O}(1/r)$ term makes its energy diverge linearly.
On the other hand, the angular components of the current $j^{\theta}$
(\ref{jth}) and the pressure $T^{\theta}_{\;\theta}$ (\ref{Tth})
have different behaviors for the leading terms. They have a power-law
decay in this case while those for the linearly-growing tachyon have
an exponential decay (\ref{logsolT}),
\begin{eqnarray}
j^{\theta}&\approx & \frac{64 n\,\tilde{{\cal
T}_{3}^2}R^{2}}{\sqrt{1-E_z^2}~|Q_{\rm F1}|} \, \frac{(\ln
r)^{2}}{r^{5}} + \cdots ,
\\
T^{\theta}_{\; \theta}&\approx & -\frac{16\tilde{{\cal
T}}_3^2}{\sqrt{1-E_z^2}~|Q_{\rm F1}|} \,\frac{1}{r^{3}} +\cdots .
\end{eqnarray}

\begin{figure}
\vspace{-15mm}
\centerline{\epsfig{figure=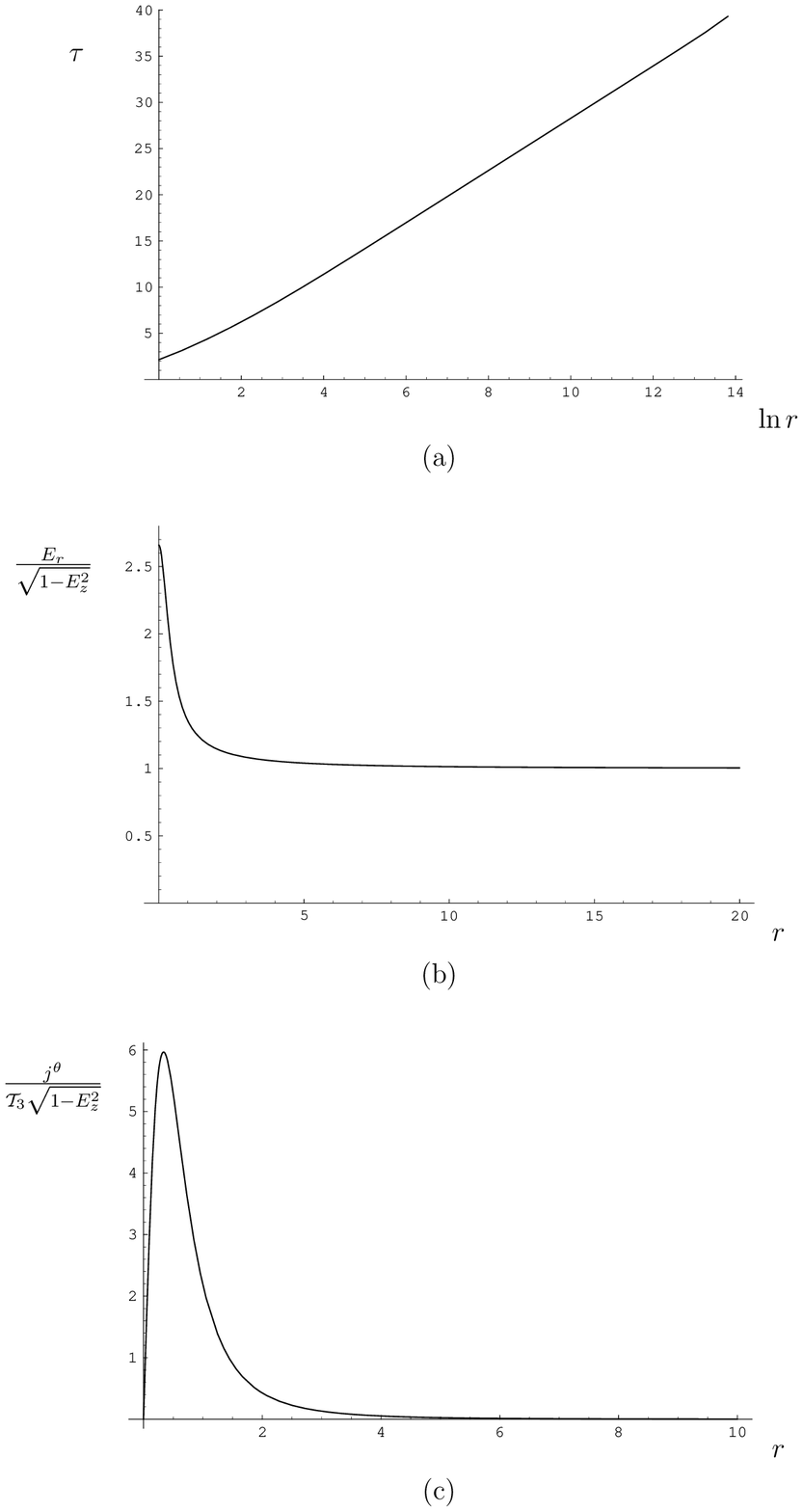,height=210mm}}
\caption{
(a) Plot of $\tau(r)$ with logarithmic behavior for large $r$, (b)
Plot of $E_{r}(r)$, and (c) Plot of $j^{\theta}$. We set $n=1$,
$Q_{{\rm F1}}/{\cal T}_{3}=2$, $R=\sqrt{2}$, and
$\tau_{0}=2.46327$. }
\label{fig3}
\end{figure}

The numerical solution for the logarithmic tachyon amplitude
connecting the origin and large $r$ is
shown in Fig.~\ref{fig3}-(a). We read $\tau_{0}$ as $\tau_{0}
=2.46327$. The profile of the radial electric field $E_{r}$
decreases monotonically from a nonzero value larger than unity
at the origin to unity at infinity as shown in Fig.~\ref{fig3}-(b).
The angular components of the current $j^{\theta}$ and
the pressure $T^{\theta}_{\theta}$ have a ring shape connecting zeros
at both boundaries, and $j^{\theta}$ is plotted in Fig.~\ref{fig3}-(c).
Since $\Pi^{z}$, $T^{t}_{t}$, $T^{r}_{r}$, and $T^{z}_{z}$ behave
in a similar way, we only draw the figure of $\Pi^{z}$
which has ${\cal O}(1/r)$ singularity at the origin and decreases
monotonically to zero as shown in Fig.~\ref{fig2}.
Both the power-law solution (\ref{solT}) and the logarithmic
solution (\ref{logsolT}) share similar shapes for $\Pi^{z}$
as are given by the solid and dashed lines in Fig.~\ref{fig2}.

The leading linear divergence in the energy of the obtained DF-string
configurations per unit length along the $z$-direction
can be read off from (\ref{ett2}) and (\ref{trin}),
\begin{equation}\label{Eg}
\frac{E}{\int dz} =\int_0^{R_{{\rm IR}}}dr r d\theta \, T_{tt}=
2\pi\sqrt{\frac{1+\tau_{\infty}^{2}}{1-E_{z}^{2}}}\, |Q_{{\rm
F1}}|R_{{\rm IR}}+({\rm finite}) .
\end{equation}
Since the divergent part is linearly proportional to the
fundamental-string charge density at the origin, a possible source
of this infra-red divergence is different from the familiar nature
of logarithmically divergent energy of the global vortex. For a
given fundamental-string charge density $Q_{{\rm F1}}$, the energy
spectrum of each solution is classified by $\tau_{\infty}$. In
that sense, the logarithmic solution with $\tau_{\infty}=0$ in
(\ref{Eg}) is the minimum energy solution of the D- or DF-strings.

If we take the limit of vanishing fundamental-string charge
density $Q_{{\rm F}1}\rightarrow 0$, the first terms proportional
to $Q_{{\rm F}1}^{2}$ in $T^{t}_{\;t}$, (\ref{tt0}) and
(\ref{ett2}) approach zero for the linearly-growing solution
(\ref{solT}). From behavior of the second $Q_{{\rm
F}1}$-independent terms in (\ref{tt0}) and (\ref{ett2}), we may
read the limit of $\delta$-function like configuration in the
limit of $\tau_{\infty}\rightarrow \infty$. This phenomenon is
consistent with the existence of singular vortex solution in the
absence of $Q_{{\rm F}1}^{2}$~\cite{Sen:2003tm}. When the
electric field $E_{z}$ approaches critical value, various
densities including $\Pi^{z}$ (\ref{Piz}), $T^{t}_{\;t}$
(\ref{Tt}), $T^{z}_{\; z}$ (\ref{Tz}), $j^{\theta}$ (\ref{jth}),
and $T^{\theta}_{\;\theta}$ diverge with the finite
fundamental-string charge density $Q_{{\rm F}1}$, but $E_{r}$
(\ref{Er}) and $T^{r}_{\; r}$ (\ref{Tr}) go to zero. These can
easily be checked also by the expanded expressions given in this
subsection. This singularity was expected from the beginning if we
see the expression of determinant (\ref{det1}) in the action
(\ref{actl}).

\begin{figure}[t]
\centerline{\epsfig{figure=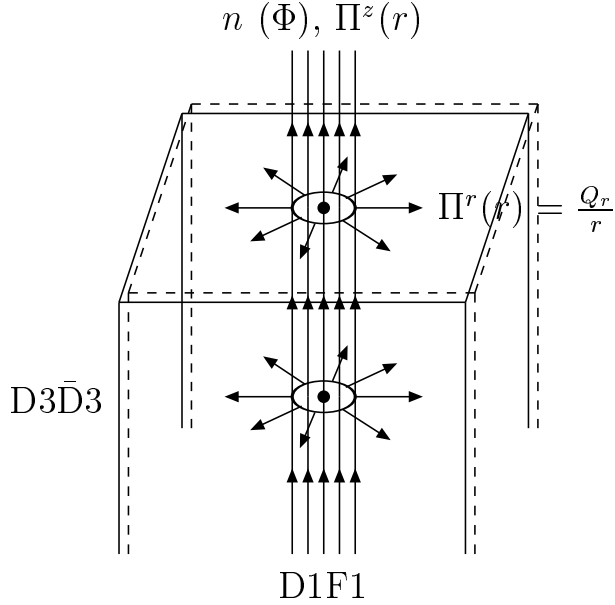,height=80mm}}
\vspace{0.2in}
\caption{D3 and ${\bar
{\rm D}}3$ with fundamental strings represented by $\Pi^{r}(r)$.
Straight DF-strings represented by $n$ and $\Pi^{z}$ are formed.}
\label{fig4}
\end{figure}

There is another coupling to the bulk RR fields given by the Wess-Zumino
term, and, for the global DF-strings from
D$3{\bar {\rm D}}3$~\cite{Kraus:2000nj,Jones:2002si,Sen:2003tm,Kennedy:1999nn},
it is
\begin{eqnarray}
S_{\rm WZ}&=& \mu ~{\rm Str}\int_{\Sigma_4}C_{\rm RR}\wedge{\rm
exp}\left(\begin{array}{cc} F^{1}-T\bar{T} & i^{3/2}~\partial T \\
-i^{3/2}~\partial\overline{T} & F^{2}-\bar{T}T
\end{array}\right)
\nonumber \\
&=&-n\mu\int_{\Sigma_4} \frac{de^{-\tau^{2}}}{dr}
\left(C_{\rm RR}^{(1)}\wedge dr\wedge d\theta +
\frac{E_{z}}{3}C_{\rm RR}^{(-1)}\wedge dt\wedge dr\wedge d\theta\wedge dz
\right),
\label{rtwz}\\
&\propto & n,
\end{eqnarray}
where $\mu$ is a real constant and the supertrace ${\rm Str}$ is
defined to be a trace with $\sigma_3$ inserted. The first term in
(\ref{rtwz}) means the charge of a D1-brane stretched along the
$z$-axis, which is proportional to the vorticity $n$. Thus the
charge density of the D1-brane per unit length is identified as
the topological charge of which current density is defined by
\begin{equation}\label{d1c}
j_{{\rm D}1}^{\mu}=\frac{{\bar T}\partial^{\mu}T-T\partial^{\mu}{\bar T}}{
4\pi i{\bar T}T}.
\end{equation}
Although the second term proportional to both $n$ and $E_{z}$ in
(\ref{rtwz}) implies an (Minkowski) instanton charge, its possible
physical meaning will be discussed in the next subsection. In
addition to the D1 charge (\ref{d1c}), the stringy object of
interest carries the charge of fundamental strings along the
$z$-axis, which is denoted by the localized electric flux
$\Pi^{z}$ (\ref{Piz}). Since the point charge $Q_{{\rm F}1}$
(\ref{Pir}) at $r=0$ is nothing but the background charge
distribution of fundamental strings coming from a transverse
direction, and is ending on a point along the $z$-axis, the
stringy object carrying the vortex charge $n$ and the localized
electric flux $\Pi^{z}$ is identified as a DF-string or a
$(p,q)$-string (composite of D1F1) from D3${\bar {\rm D}}3$ system
with fundamental strings. What we obtained is summarized
schematically in Fig.~\ref{fig4}. If the early Universe involved a
D3${\bar {\rm D}}3$, the obtained DF-string can remain as a cosmic
fossil named as the cosmic global DF-string.

\subsection{Local DF-strings}
When the gauge field $C_{\mu}$ (\ref{anc}) is turned on, the character of
DF-strings becomes local. In usual local field theories, e.g., the
Abelian-Higgs model, a role of the gauge field is to make energy of the
local vortex (energy density of the vortex-string per unit length along the
string) finite by trimming the logarithmically-divergent energy of
the global vortex~\cite{VS}. This phenomenon was not observed in D-vortices
from D2${\bar {\rm D}}$2 system with fundamental strings;
the energy of the D-vortex is linearly-divergent, but its source is
fundamental string charges~\cite{Kim:2005tw}.
Although this sort of energy-trimming seems unlikely also for local
DF-strings of our interest, we investigate the existence and the property of
local DF-strings in this subsection.

Since the inclusion of the gauge field $C_{\theta}$ (\ref{anc}) requires
the analysis of the coupled equations (\ref{Teq})--(\ref{Ceq}),
we need boundary conditions for the gauge field in
addition to those for the tachyon (\ref{bd}),
\begin{equation}\label{cbd}
C_{\theta}(0)=0,\qquad C_{\theta}(\infty)=\frac{n}{2}.
\end{equation}
From now on, we examine the differential equations (\ref{Teq})--(\ref{Ceq})
and the expressions (\ref{Er}) and (\ref{Piz}) for $E_{r}$ and $\Pi^{z}$,
and find local DF-string
solutions satisfying the boundary conditions (\ref{bd}) and (\ref{cbd}).

Near the origin, the power-series expansion of $\tau(r)$ and $C_{\theta}(r)$
for DF-string solutions gives
\begin{equation}\label{tao}
\tau(r)\approx
\left\{
\begin{array}{ll}
\tau_0 r \left[1 -\frac{{\cal T}_3^2(1+\tau_0^2)^2}{5 Q_{\rm F1}^2R^2}r^4
+ \cdots\right], & n=1 \\
\tau_{0}r\left[1+\frac{(n^{2}-1)\alpha}{6} r^2+ \cdots\right], & n \ge 2
\end{array}
\right.
\end{equation}
\begin{equation}\label{cto}
C_{\theta}(r)\approx
C_{0}r^{3}\left[1-\frac{3+\tau_0^2 n^2(5-2n^2)}{10(1+\tau_0^2n^2)}\alpha r^2+
\cdots \right], \qquad  n\ge 1
\end{equation}
where $\alpha$ is
\begin{eqnarray}\label{alp}
\alpha=\frac{1}{(1+\tau_0^2n^2)^2}\left[\frac{4{\cal
T }_3^2}{Q_{\rm F1}^2}(1+\tau_0^2)(1+\tau_0^2n^2)^2-9C_0^2\right] .
\end{eqnarray}
For the local DF-strings with unit vorticity, the increasing rate of the
tachyon amplitude (\ref{tao}) is not affected by $C_{0}$ up to the
second order. On the other hand,
the signature in front of $9C_{0}^{2}$ in (\ref{alp}) is opposite
to that of the first term which is proportional to
${\cal T}_{3}^{2}/Q_{{\rm F}1}^{2}$,
so the increasing rate of the tachyon amplitude
(\ref{tao}) becomes smaller for local DF-strings.

Inserting the expansions
(\ref{tao})--(\ref{cto}) into the radial electric field $E_{r}$ (\ref{Er})
and the fundamental-string charge density $\Pi^{z}$ (\ref{Piz}), we have
a nonzero value, $(1-E_z^2)(1+\tau_{0}^2)$, for $E_{r}$ and
singular $|\Pi^{z}|$ at the origin
\begin{eqnarray}
E_{r}^{2}(r)&\approx & (1-E_z^2)(1+\tau_{0}^2)(1-\alpha r^2+
\cdots), \label{ero}\\
\Pi^{z2} &\approx &
\frac{E_z^2(1+\tau_0^2)}{(1-E_z^2)}\left(\frac{Q_{\rm
F1}}{r}\right)^2\left( 1 - \beta r^{2}+ \cdots \right),
\label{piz0}
\end{eqnarray}
where $\beta$ is
\begin{eqnarray}
\beta =- \frac{1}{(1+\tau_0^2n^2)^2}\left[
\frac{4{\cal T
}_3^2}{Q_{\rm F1}^2}(1+\tau_0^2(2n^2-1))(1+\tau_0^2n^2 )^2 + 9C_0^2
\right].
\end{eqnarray}
The current density (\ref{jth}) again increases from zero
\begin{eqnarray}
j^{\theta}\approx
\sqrt{\frac{1-E_z^2}{1+\tau_0^2}}~\tau_0^2n|Q_{\rm
F1}|\alpha r + \cdots ,
\end{eqnarray}
and nonvanishing components of the energy-momentum density
(\ref{Tt})--(\ref{Tz}) show the following behavior which is similar to
the case of global DF-strings (\ref{tt0})--(\ref{tz0})
\begin{eqnarray}
T^{t}_{\;t}&\approx &
-\sqrt{\frac{1+\tau_{0}^2}{1-E_z^2}}~\frac{|Q_{\rm
F1}|}{r}\left(1-\frac{\beta}{2}r^{2}+ \cdots \right),
\\
T^{r}_{\;r}&\approx &
-\sqrt{\frac{1-E_z^2}{1+\tau_{0}^2}}~\frac{|Q_{\rm F1}|}{r}
\left( 1 +\frac{\alpha}{2}r^{2}+ \cdots \right),
\\
T^{\theta}_{\;\theta}&\approx &
-\sqrt{\frac{1-E_z^2}{1+\tau_0^2}}\, |Q_{\rm F1}|\alpha r + \cdots
,
\\
T^z_{\;z}&\approx & -E_z^2
\sqrt{\frac{1+\tau_0^2}{1-E_z^2}}~\frac{|Q_{\rm
F1}|}{r}\left\{1-\left[\frac{\alpha}{2}-\frac{4(1+\tau_{0}^{2}n^{2})}{E_{z}^{2}
}\frac{{\cal T}_{3}^{2}}{Q_{{\rm F}1}^{2}}
\right]r^{2}+\cdots\right\}.
\end{eqnarray}

The near-origin behavior of the local DF-string solutions is parameterized
smoothly by $\tau_{0}$ in (\ref{tao}) and $C_{0}$ in (\ref{cto}).
At asymptotic region, the tachyon amplitude of every local DF-string
will be proven to approach
universally the vacuum, but the approaching behavior is
sorted into two; (i) linear divergence  $\tau \sim \tau_\infty r$
and (ii) logarithmic divergence $\tau \sim \tau_\infty \ln r$, as were
for the global DF-strings in the previous subsection.
We analyze the DF-string solutions with the linearly-divergent $\tau$ and
the logarithmically-divergent $\tau$ separately in what follows.

{\bf (i) \underline{$\tau\sim \tau_\infty r$ solutions}}:
If we examine carefully the coupled differential equations
(\ref{Teq})--(\ref{Ceq}), the leading asymptotic behavior of the tachyon
amplitude is either linearly-growing or logarithmically-growing.
First, we consider the linearly-growing case. The subleading term of
the tachyon amplitude is decreasing exponentially
\begin{eqnarray}\label{tan}
\tau(r)&\approx& \tau_{\infty}r +\delta -\frac{4{\cal
T}_3^2R(1+\tau_\infty^2)}{\tau_\infty^2Q_{\rm
F1}^2}~r^2e^{-2\frac{\tau_\infty r+\delta }{R}}+\cdots ,
\end{eqnarray}
where $\tau_{\infty}$ and $\delta$ are undetermined constants which are
governed by the behavior near the origin.
For the gauge field $C_{\theta}$, we consider small $\delta C_\theta$
at the asymptotic region,
\begin{eqnarray}\label{ctn}
C_{\theta}(r)&\approx& \frac{n}{2} + \delta C_\theta .
\end{eqnarray}
Substituting this into the equation for the gauge field, we obtain a linear
equation,
\begin{eqnarray}\label{Ne}
M(t)\frac{d^2\delta C_\theta}{dt^2}
=-\frac{d}{d\delta C_\theta}U(\delta C_\theta),\qquad
t=\kappa r^3 , \quad \left(
\kappa=\frac{4{\cal T}_p\tau_\infty}{3|Q_{\rm F1}|}
\sqrt{1+\tau_\infty^2}e^{-\delta/R}
\right),
\end{eqnarray}
where $M(t)=e^{2\tau_\infty t^{1/3}/(R\kappa^{1/3})}$ and
$U=-(\delta C_\theta)^2/2$.
If we identify $\delta C_\theta$ as a one-dimensional position of a hypothetical
particle, Eq.~(\ref{Ne}) is interpreted as a Newtonian equation with
a variable mass $M(t)$
and a conservative potential $U(\delta C_\theta)$. The nontrivial analytic
solution satisfying
$\delta C_\theta (r=\infty)=0$ is not known yet, but the
existence of such solution
can easily be read from the properties of $U(\delta C_\theta)$;
max[$U(\delta C_\theta)$]=0 at $\delta C_\theta=0$ and
min[$U(\delta C_\theta)]=-\infty$ at $\delta C_\theta=\pm\infty$.

With the asymptotic behavior of the tachyon amplitude (\ref{tan}) and
the gauge field (\ref{ctn}), the radial electric field approaches
its boundary value exponentially,
\begin{eqnarray}\label{ern}
E_{r}^{2}(r)&\approx& (1-E_z^2)(1+\tau_{\infty}^2)
\left(1-\frac{16{\cal T}_3^2R}{\tau_\infty Q_{\rm
F1}^2}re^{-2\frac{\tau_\infty r+\delta }{R}}+\cdots \right),
\end{eqnarray}
and the U(1) current and the angular component of the energy-momentum tensor
decay exponentially to zero,
\begin{eqnarray}
j^{\theta}&\approx & -32{\tilde{\cal
T}}_3^2\tau_\infty^2\sqrt{\frac{1+\tau_\infty^2}{1-E_z^2}}~\frac{r}{|Q_{\rm
F1}|}e^{-2\frac{\tau_\infty r +\delta}{R}}\delta C_{\theta} ,
\label{jtil} \\
T^{\theta}_{\;\theta}&\approx &  -16\tilde{{\cal
T}}_3^2\sqrt{\frac{1+\tau_\infty^2}{1-E_z^2}}~\frac{r}{|Q_{\rm
F1}|}e^{-2\frac{\tau_\infty r+\delta }{R}}
+ \cdots .
\label{thhil}
\end{eqnarray}
As it was the case of global DF-strings in the previous
subsection, the leading terms of $\Pi^{z}$, $T^{t}_{\;t}$,
$T^{r}_{\;r}$, and $T^z_{\;z}$ are commonly proportional to
$Q_{\rm F1}/r$ but the subleading terms decrease exponentially,
\begin{eqnarray}
\Pi^{z2} &\approx&
\frac{E_z^2(1+\tau_\infty^2)}{1-E_z^2}\left(\frac{Q_{\rm
F1}}{r}\right)^2\left(1+\frac{32{\cal T }_3^2}{Q_{\rm F1
}^2}r^2e^{-2\frac{\tau_\infty r+\delta }{R}}+\cdots \right) ,
\label{pizi}\\
T^{t}_{\;t}&\approx &
-\sqrt{\frac{1+\tau_{\infty}^2}{1-E_z^2}}~\frac{|Q_{\rm F1}|}{r}
\left(1+\frac{16{\cal T}_3^2}{Q_{\rm
F1}^2}~r^2e^{-2\frac{\tau_\infty r+\delta }{R}}+\cdots \right),
\label{etil}\\
T^{r}_{\;r}&\approx &
-\sqrt{\frac{1-E_z^2}{1+\tau_{\infty}^2}}~\frac{|Q_{\rm F1}|}{r}
\left(1+\frac{8{\cal T}_3^2R}{\tau_\infty Q_{\rm F1}^2}
re^{-2\frac{\tau_\infty r+\delta }{R}} + \cdots \right),
\label{eril}\\
T^z_{\;z}&\approx&
-\sqrt{\frac{1+\tau_\infty^2}{1-E_z^2}}~\frac{|Q_{\rm
F1}|}{r}\left(E_z^2+\frac{16{\cal T}_3^2}{Q_{\rm
F1}^2}r^2e^{-2\frac{\tau_\infty r+\delta }{R}} +\cdots \right) .
\end{eqnarray}
Note that the leading long-range terms of local DF-strings are the
same as those of global DF-strings, and that the limiting behaviors for the
large, or small
$Q_{{\rm F}1}^{2}$, and for the critical electric field $|E_{z}|\rightarrow 1$,
are also the same.

\begin{figure}[t]
\centerline{\epsfig{figure=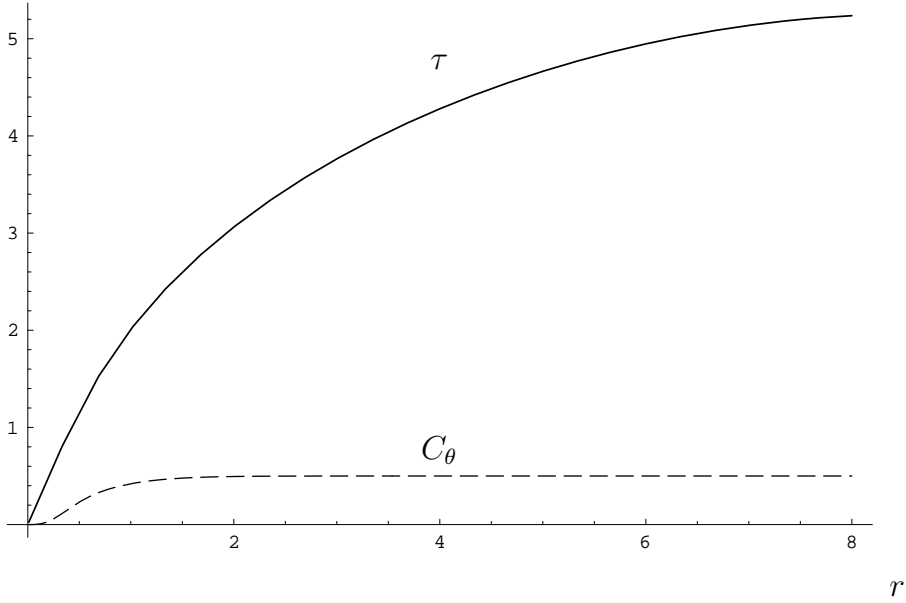,height=80mm}}
\caption{For $n=1$ and $R=\sqrt{2}$,
profile of the tachyon amplitude $\tau(r)$ by the solid line
($\tau_{0}=2.4489$) and that of the gauge field $C_{\theta}(r)$ by the
dashed line ($C_{0}=4.5733$).}
\label{fig5}
\end{figure}

{\bf (ii) \underline{$\tau\sim \tau_\infty \ln r$ solutions}}:
Lastly let us discuss logarithmic $\tau$-solution of the local
DF-string. For sufficiently large $r$, the leading logarithmic
term has the same coefficient with the global DF-string
(\ref{logsolT}), but the subleading term does not contain
logarithmic term;
\begin{eqnarray}
\tau(r)&\approx& 2R ~{\rm ln}r -\frac{4R^3}{r^2} +\cdots .
\end{eqnarray}
As we observed in the global string case, the first subleading term of
$E_{r}$ in (\ref{loer}) is governed only by the leading term of the tachyon
amplitude in (\ref{logsolT}). Therefore, the expansion for the
logarithmic $\tau$-solution is the same for the local string.
If we try to get the asymptotic solution of the gauge field $C_{\theta}$
similarly to the case of linearly-growing solution (\ref{ctn}), the
linearized equation for $\delta C_\theta$ is
\begin{equation}
\frac{d}{dr}\left(\frac{\delta C_\theta'}{r^2}\right)
\approx 32R^4\frac{({\rm ln}r)^2}{r^4}\delta C_\theta .
\end{equation}
We find a solution which decreases rapidly to zero,
\begin{equation}
\delta C_\theta (r)\approx C_{1}\frac{r^{\frac{3}{2}}}{\sqrt{\ln r}}\,
{\rm WhittakerW}\left(-\frac{9}{128}\frac{\sqrt{2}}{R},\, \frac{1}{4},\,
4\sqrt{2}R(\ln r)^{2}\right),
\end{equation}
where $C_{1}$ is an integration constant which is to be set by the
boundary conditions at the asymptotic.
As shown in Fig.~\ref{fig5}, the gauge field $C_{\theta}$
(the dashed line) reaches its boundary value $n/2$ rapidly and
the tachyon amplitude $\tau$ (the solid line) grows logarithmically.
Due to the mixture of slowly-growing and rapidly-growing functional
behaviors for single numerical analysis, the results of numerical work
need further improvement for this logarithmic case.

This rapidly-decreasing behavior of the gauge field $C_{\theta}$
affects that of the U(1) current (\ref{jth}), but does not appear in
the radial pressure (\ref{Tr}) up to the leading order,
\begin{eqnarray}
j^{\theta}&\approx & -\frac{128\,\tilde{{\cal
T}_{3}^2}R^{2}}{\sqrt{1-E_z^2}~|Q_{\rm F1}|} \,
\frac{(\ln r)^{2}}{r^{5}} \delta C_\theta +\cdots  , \\
T^{\theta}_{\; \theta}&\approx & -\frac{16\tilde{{\cal
T}}_3^2}{\sqrt{1-E_z^2}~|Q_{\rm F1}|}
\,\frac{1}{r^{3}}+\cdots .
\end{eqnarray}
Similar to $E_{r}$, all the components of the energy-momentum tensor and
the $z$-component of the electric flux density for the local DF-string
have the same functional forms with those of the corresponding components
of the global DF-string up to the second order
terms. Therefore, the leading energy density of the local DF-string for the
logarithmically-growing tachyon amplitude shares that of
the local DF-string for the linearly-growing tachyon amplitude and
that of the global strings. It means that the discussion on the energy
of the global DF-strings can also be applied to that of the local DF-strings;
the role of the gauge field $C_{\theta}$ in localizing the energy
of the string is negligible, which is very different from that in the case of
Nielsen-Olesen vortex-string in the Abelian-Higgs model.

Due to the gauge field $C_{\theta}$, the Wess-Zumino term
of D$3{\bar {\rm D}}3$ describing coupling to the bulk RR fields
becomes slightly different from that of the global
DF-strings~\cite{Kraus:2000nj,Jones:2002si,Sen:2003tm,Kennedy:1999nn},
\begin{eqnarray}
S_{\rm WZ}&=& \mu ~{\rm Str}\int_{\Sigma_4}C_{\rm RR}\wedge{\rm
exp}\left(\begin{array}{cc} F^{1}-T\bar{T} & i^{3/2}~D T \\
-i^{3/2}~\overline{DT} & F^{2}-\bar{T}T
\end{array}\right)
\nonumber \\
&=&2\mu\int_{\Sigma_4} \left\{\frac{d}{dr}\left[e^{-\tau^{2}}\left(C_{\theta}-
\frac{n}{2}\right)\right]
\left(C_{\rm RR}^{(1)}+\frac{E_{z}}{3}C_{\rm RR}^{(-1)}\wedge dt\wedge dz
\right)\wedge dr\wedge d\theta
\right. \nonumber\\
&&\hspace{15mm} \left. -\frac{1}{3}E_{z}C_{r\theta}
C_{\rm RR}^{(-1)}\wedge dt\wedge dr\wedge d\theta\wedge dz
\right\}.
\label{rtz}
\end{eqnarray}
The term of $C_{\rm RR}^{(1)}$ coupling is proportional to the vorticity $n$
so that the local DF-string carries the quantized magnetic flux as
a D1 charge density per unit length along the $z$-axis,
\begin{eqnarray}\label{flu}
\Phi=\int_{0}^{\infty} dr\int_{0}^{2\pi}d\theta \,
C_{r\theta} =\pi n.
\end{eqnarray}
Note that the second term in (\ref{rtz}) tells us that $C_{\rm RR}^{(-1)}$
coupling is nothing but an axion coupling.
Since we have additionally the fundamental-string charge density $\Pi^{z}$
localized along the string direction (the $z$-axis in our case), the obtained
stringy objects are local DF-strings, or local $(p,q)$-strings from
D3${\bar {\rm D}}$3 system with fundamental strings.

\setcounter{equation}{0}
\section{Conclusions}
The system of D3${\bar {\rm D}}3$ with fundamental strings has
been considered in the coincidence limit of a brane and an
anti-brane. In the scheme of effective field theory, it is
described by the DBI type effective action of a complex tachyon
field and U(1)$\times$U(1) gauge fields. The runaway tachyon
potential has U(1) vacua at an infinite tachyon amplitude, which
supports topological vortex-strings of codimension-two.
Specifically, we study straight string solutions by examining
field equations.

The topological charge of the string represented by vorticity is
interpreted as the RR-charge density of D-string (D1-brane).  (See
the circles in Fig.~\ref{fig4}.) Introduction of the radial DBI
electric field coupled nonminimally to the tachyon is
indispensable to obtain a thick D-string, which implies that
background fundamental strings live in an extra-dimension with a
fluid form and end at each vortex-string origin. (See the radial
arrows in Fig.~\ref{fig4}.)

According to asymptotic behavior of the tachyon amplitude at
infinity, we obtained either linearly-growing tachyon
configurations, or a newly-found lograrithmically-growing tachyon
configuration which represents the minimum energy configuration.
We additionally turn on the constant DBI electric field parallel
to the string. Then its conjugate momentum density is localized
along the string. (See the straight arrows along the $z$-direction
in Fig.~\ref{fig4}.) This confined electric flux density tells us
that the stringy object of interest carries a fundamental string
charge density, so we find it a DF-string (or a $(p,q)$-string).
Lastly the nonvanishing gauge field coupled minimally to the
tachyon replaces the global DF-string by a local DF-string
carrying a quantized magnetic flux density as a D1 charge density.

Now that we have global and local, D- and DF-strings as soliton
solutions in the context of effective field theory~\cite{Kim:2005tw},
the future tasks to construct a viable cosmic superstring become more
tractable. Dynamical issues~\cite{Dvali:2003zj,Copeland:2003bj,Leblond:2004uc}
involve (i) head-on collision of two
D-vortices for checking the intercommuting (reconnecting) property
of two D-strings, (ii) collision of two DF-strings leading to a
tree structure composed of a pair of trilinear vertices, which is
to form a cosmic DF-string network~\cite{Hashimoto:2002xe},
and (iii) the stability of long macroscopic D- and DF-strings.

On cosmological aspects, we would gravitate the obtained static
stringy defects and see the resultant spacetime structure. This
may provide a basis to tackle the possibility of observing its
effect astrophysically including viable density fluctuations and
quintessence~\cite{Cho:1998jk}. Inclusion of time-dependence is
also important to understand how the D- and DF-strings are
generated, and whether or not they survive during the inflationary
era induced by the separation of D3 and ${\bar {\rm D}}3$. To
lower the scale from the fundamental string scale to a scale to
pass observational tests such as the cosmic microwave background,
the pulsar timing, and the gravitational radiation, it is also
intriguing to take into account the D- and DF-strings obtained in
the background of various string (inspired)
models~\cite{Sarangi:2002yt}.

\section*{Acknowledgments}
We would like to thank Jungjai Lee, Sangmin Lee, and Jin Hur for
helpful discussions. This work is the result of research
activities (Astrophysical Research Center for the Structure and
Evolution of the Cosmos (ARCSEC)) supported by Korea Science $\&$
Engineering Foundation (Y.K.), and the BK 21 project of the
Ministry of Education and Human Resources Development, Korea
(I.C.).


\begin{thebibliography}{100}

\bibitem{Kraus:2000nj}
P.~Kraus and F.~Larsen,
``Boundary string field theory of the DD-bar system,''
Phys.\ Rev.\ D {\bf 63}, 106004 (2001)
[arXiv:hep-th/0012198];

T.~Takayanagi, S.~Terashima and T.~Uesugi,
``Brane-antibrane action from boundary string field theory,''
JHEP {\bf 0103}, 019 (2001)
[arXiv:hep-th/0012210].

\bibitem{Jones:2002si}
N.~T.~Jones and S.~H.~H.~Tye,
``An improved brane anti-brane action from boundary superstring field theory
and multi-vortex solutions,''
JHEP {\bf 0301}, 012 (2003)
[arXiv:hep-th/0211180].

\bibitem{Sen:2003tm}
A.~Sen,
``Dirac-Born-Infeld action on the tachyon kink and vortex,''
Phys.\ Rev.\ D {\bf 68}, 066008 (2003)
[arXiv:hep-th/0303057].

\bibitem{Dvali:1998pa}
G.~R.~Dvali and S.~H.~H.~Tye,
``Brane inflation,''
Phys.\ Lett.\ B {\bf 450}, 72 (1999)
[arXiv:hep-ph/9812483];

G.~R.~Dvali, Q.~Shafi and S.~Solganik,
``D-brane inflation,''
arXiv:hep-th/0105203;

C.~P.~Burgess, M.~Majumdar, D.~Nolte, F.~Quevedo, G.~Rajesh and R.~J.~Zhang,
``The inflationary brane-antibrane universe,''
JHEP {\bf 0107}, 047 (2001)
[arXiv:hep-th/0105204];

J.~Garcia-Bellido, R.~Rabadan and F.~Zamora,
``Inflationary scenarios from branes at angles,''
JHEP {\bf 0201}, 036 (2002)
[arXiv:hep-th/0112147];

S.~H.~S.~Alexander,
``Inflation from D - anti-D brane annihilation,''
Phys.\ Rev.\ D {\bf 65}, 023507 (2002)
[arXiv:hep-th/0105032];

E.~Halyo,
``Inflation from rotation,''
arXiv:hep-ph/0105341;

G.~Shiu and S.~H.~H.~Tye,
``Some aspects of brane inflation,''
Phys.\ Lett.\ B {\bf 516}, 421 (2001)
[arXiv:hep-th/0106274];

C.~Herdeiro, S.~Hirano and R.~Kallosh,
``String theory and hybrid inflation / acceleration,''
JHEP {\bf 0112}, 027 (2001)
[arXiv:hep-th/0110271];

B.~s.~Kyae and Q.~Shafi,
``Branes and inflationary cosmology,''
Phys.\ Lett.\ B {\bf 526}, 379 (2002)
[arXiv:hep-ph/0111101];

R.~Blumenhagen, B.~Kors, D.~Lust and T.~Ott,
``Hybrid inflation in intersecting brane worlds,''
Nucl.\ Phys.\ B {\bf 641}, 235 (2002)
[arXiv:hep-th/0202124];

N.~Jones, H.~Stoica and S.~H.~H.~Tye,
``Brane interaction as the origin of inflation,''
JHEP {\bf 0207}, 051 (2002)
[arXiv:hep-th/0203163].

\bibitem{VS} A. Vilenkin and E.~P.~S.~Shellard,
{\it Cosmic Strings and Other Topological Defects},
(Cambridge University Press, 1984).

\bibitem{Kibble:2004hq}
T.~W.~B.~Kibble,
``Cosmic strings reborn?,''
arXiv:astro-ph/0410073;

A.~Vilenkin,
``Cosmic strings: Progress and problems,''
arXiv:hep-th/0508135.

\bibitem{Dvali:2003zj}
G.~Dvali and A.~Vilenkin,
``Formation and evolution of cosmic D-strings,''
JCAP {\bf 0403}, 010 (2004)
[arXiv:hep-th/0312007].

\bibitem{Copeland:2003bj}
E.~J.~Copeland, R.~C.~Myers and J.~Polchinski,
``Cosmic F- and D-strings,''
JHEP {\bf 0406}, 013 (2004)
[arXiv:hep-th/0312067].

\bibitem{Leblond:2004uc}
L.~Leblond and S.~H.~H.~Tye,
``Stability of D1-strings inside a D3-brane,''
JHEP {\bf 0403}, 055 (2004)
[arXiv:hep-th/0402072];

T.~Matsuda,
``String production after angled brane inflation,''
Phys.\ Rev.\ D {\bf 70}, 023502 (2004)
[arXiv:hep-ph/0403092];

M.~G.~Jackson, N.~T.~Jones and J.~Polchinski,
``Collisions of cosmic F- and D-strings,''
arXiv:hep-th/0405229;

K.~Dasgupta, J.~P.~Hsu, R.~Kallosh, A.~Linde and M.~Zagermann,
``D3/D7 brane inflation and semilocal strings,''
JHEP {\bf 0408}, 030 (2004)
[arXiv:hep-th/0405247];

A.~Achucarro and J.~Urrestilla,
``F-term strings in the Bogomolnyi limit are also BPS states,''
JHEP {\bf 0408}, 050 (2004)
[arXiv:hep-th/0407193];

T.~Damour and A.~Vilenkin,
``Gravitational radiation from cosmic (super)strings: Bursts, stochastic
background, and observational windows,''
Phys.\ Rev.\ D {\bf 71}, 063510 (2005)
[arXiv:hep-th/0410222];

N.~Barnaby, A.~Berndsen, J.~M.~Cline and H.~Stoica,
``Overproduction of cosmic superstrings,''
JHEP {\bf 0506}, 075 (2005)
[arXiv:hep-th/0412095];

H.~Firouzjahi and S.~H.~Tye,
``Brane inflation and cosmic string tension in superstring theory,''
JCAP {\bf 0503}, 009 (2005)
[arXiv:hep-th/0501099];

S.~H.~Tye, I.~Wasserman and M.~Wyman,
``Scaling of multi-tension cosmic superstring networks,''
Phys.\ Rev.\ D {\bf 71}, 103508 (2005)
[Erratum-ibid.\ D {\bf 71}, 129906 (2005)]
[arXiv:astro-ph/0503506];

E.~J.~Copeland and P.~M.~Saffin,
``On the evolution of cosmic-superstring networks,''
arXiv:hep-th/0505110;
%

A.~C.~Davis and K.~Dimopoulos,
``Cosmic superstrings and primordial magnetogenesis,''
Phys.\ Rev.\ D {\bf 72}, 043517 (2005)
[arXiv:hep-ph/0505242];

P.~M.~Saffin,
``A practical model for cosmic (p,q) superstrings,''
JHEP {\bf 0509}, 011 (2005)
[arXiv:hep-th/0506138];

E.~I.~Buchbinder,
``On Open Membranes, Cosmic Strings and Moduli Stabilization,''
Nucl.\ Phys.\ B {\bf 728}, 207 (2005)
[arXiv:hep-th/0507164];

B.~Shlaer and M.~Wyman,
``Cosmic superstring gravitational lensing phenomena: Predictions for
networks of (p,q) strings,''
arXiv:hep-th/0509177.

\bibitem{Blanco-Pillado:2005xx}
J.~J.~Blanco-Pillado, G.~Dvali and M.~Redi, ``Cosmic D-strings as
axionic D-term strings,'' arXiv:hep-th/0505172;

\bibitem{Kim:2005tw}
Y.~Kim, B.~Kyae and J.~ Lee,
``Global and local D-vortices,''
JHEP {\bf 0510}, 002 (2005)
[arXiv:hep-th/0508027].

\bibitem{Polchinski:2004ia}
For a review, see
J.~Polchinski,
``Introduction to cosmic F- and D-strings,''
arXiv:hep-th/0412244.

\bibitem{Witten:1985fp}
E.~Witten,
``Cosmic Superstrings,''
Phys.\ Lett.\ B {\bf 153}, 243 (1985).

\bibitem{Jones:2003ae}
N.~T.~Jones, L.~Leblond and S.~H.~H.~Tye,
``Adding a brane to the brane anti-brane action in BSFT,''
JHEP {\bf 0310}, 002 (2003)
[arXiv:hep-th/0307086].

\bibitem{Sen:1999mg}
A.~Sen
``Non-BPS states and branes in string theory,''
arXiv:hep-th/9904207.

\bibitem{Sen:1999xm}
A.~Sen,
``Universality of the tachyon potential,''
JHEP {\bf 9912}, 027 (1999)
[arXiv:hep-th/9911116].

\bibitem{Sen:2002an}
A.~Sen, ``Field theory of tachyon matter,'' Mod.\ Phys.\ Lett.\ A
{\bf 17}, 1797 (2002) [arXiv:hep-th/0204143].

\bibitem{Kutasov:2003er}
D.~Kutasov and V.~Niarchos, ``Tachyon effective actions in open
string theory,'' Nucl.\ Phys.\ B {\bf 666}, 56 (2003)
[arXiv:hep-th/0304045].

\bibitem{Buchel:2002tj}
A.~Buchel, P.~Langfelder and J.~Walcher, ``Does the tachyon
matter?,'' Annals Phys.\  {\bf 302}, 78 (2002)
[arXiv:hep-th/0207235];

C.~Kim, H.~B.~Kim, Y.~Kim and O-K.~Kwon, ``Electromagnetic string
fluid in rolling tachyon,'' JHEP {\bf 0303}, 008 (2003)
[arXiv:hep-th/0301076];

F.~Leblond and A.~W.~Peet, ``SD-brane gravity fields and rolling
tachyons,'' JHEP {\bf 0304}, 048 (2003) [arXiv:hep-th/0303035].

\bibitem{Kim:2003in}
C.~Kim, Y.~Kim and C.~O.~Lee,
``Tachyon kinks,''
JHEP {\bf 0305}, 020 (2003)
[arXiv:hep-th/0304180];

A.~Sen,
``Open and closed strings from unstable D-branes,''
Phys.\ Rev.\ D {\bf 68}, 106003 (2003)
[arXiv:hep-th/0305011];

C.~Kim, Y.~Kim, O-K.~Kwon and C.~O.~Lee,
``Tachyon kinks on unstable Dp-branes,''
JHEP {\bf 0311}, 034 (2003)
[arXiv:hep-th/0305092].

\bibitem{Kennedy:1999nn}
C.~Kennedy and A.~Wilkins,
``Ramond-Ramond couplings on brane-antibrane systems,''
Phys.\ Lett.\ B {\bf 464}, 206 (1999)
[arXiv:hep-th/9905195];

\bibitem{Hashimoto:2002xe}
K.~Hashimoto and S.~Nagaoka,
``Realization of brane descent relations in effective theories,''
Phys.\ Rev.\ D {\bf 66}, 026001 (2002)
[arXiv:hep-th/0202079].

\bibitem{Cho:1998jk}
I.~Cho and A.~Vilenkin,
``Vacuum defects without a vacuum,''
Phys.\ Rev.\ D {\bf 59}, 021701 (1999)
[arXiv:hep-th/9808090];

I.~Cho and A.~Vilenkin,
``Gravitational field of vacuumless defects,''
Phys.\ Rev.\ D {\bf 59}, 063510 (1999)
[arXiv:gr-qc/9810049].

\bibitem{Sarangi:2002yt}
S.~Sarangi and S.~H.~H.~Tye,
``Cosmic string production towards the end of brane inflation,''
Phys.\ Lett.\ B {\bf 536}, 185 (2002)
[arXiv:hep-th/0204074];

S.~Kachru, R.~Kallosh, A.~Linde, J.~Maldacena, L.~McAllister and S.~P.~Trivedi,
``Towards inflation in string theory,''
JCAP {\bf 0310}, 013 (2003)
[arXiv:hep-th/0308055];

K.~Becker, M.~Becker and A.~Krause,
``Heterotic cosmic strings,''
arXiv:hep-th/0510066.


\end{thebibliography}
\end{document}